# Feasibility of space-based measurement-device-independent quantum key distribution

Xingyu Wang[1, 2†], Chen Dong[1*†], Shanghong Zhao[2], Yong Liu[1], Xiaowen Liu[1], Haonan Zhu[2]

[1] *Information and Communication College, National University of Defense Technology, Xi'an, 710006, China.*

[2] *School of Information and Navigation, Air Force Engineering University, Xi'an 710077, China.*

**Abstract:** The measurement-device-independent (MDI) QKD is considered to be an alternative to overcome the currently trusted satellite paradigm. However, the feasibility of the space-based MDI-QKD remains unclear in terms of the factors: the high-loss uplink between a ground station and a satellite, the limited duration when two ground stations are simultaneously visible, as well as the rigorous requirements for the two-photon interference when performing the Bell-state Measurement (BSM). In this paper, we present a feasibility assessment of space-based MDI-QKD based on the Micius satellite. Integrated with the orbital dynamics model and atmosphere channel model, a framework is presented to explore the whole parameters space including orbit height, elevation angle, apertures of transceiver and atmospheric turbulence intensity to give the considerations for improving key rates and subsequently provide a relevant parameter tradeoff for the implementation of space-based MDI-QKD. We further investigate the heart of MDI-QKD, the two-photon interference considerations such as the frequency calibration and time synchronization technology against Doppler shift, and the way of performing the intensity optimization method in the dynamic and asymmetric channels. Our work can be used as a pathfinder to support decisions involving as the selection of the future quantum communication satellite missions.

## I. INTRODUCTION

Quantum key distribution (QKD) is a family of protocols which enables two remote parties to securely exchange cryptographic keys. The rapid development of the ground-based QKD schemes and implementations makes it realizable that the communication range reaching as far as 509 km with modern technology in the optical fiber channel [1]. Despite these developments, there is still a large gap in the goals of achieving a global quantum key distribution network. One promising solution is to use satellite terminals to distribute keys between earth stations; especially a milestone experiment has successfully demonstrated the feasibility of space-based QKD using the quantum experimental science satellite (QUESS) Micius [2-4], which has sparked a lot of interest and research in space-based quantum communication in recent years.

So far many space-based QKD schemes follow the trusted nodes approach, which envisions the satellite as a flying trusted-node for communication between locations on the ground. A recent experiment shows the feasibility of the entanglement-based QKD using a double-downlink to overcome the trusted nodes paradigm currently used for satellite based QKD [5]. Whereas, the MDI-QKD [6-7], which is a time-reversed Einstein-Podolsky-Rosen (EPR) protocol [8], gives a possibility of space-based QKD without the need for trusted satellites owing to the intrinsic interrelation of the two schemes. Inspired by the configuration of the entanglement-based QKD with the Micius satellite, we suggest that the space-based MDI-QKD providing secure communications for two simultaneously visible optical ground stations (OGS).

Achievements of space-based QKD experiments have been made in [9], which provide the basis for performing space-based MDI-QKD based on the satellite. For instance, an uplink loss under the detailed experimental parameters [10] gives a reference for modeling and optimization, where a reliable transfer of quantum state is achieved even passing through the uplink. In addition, the possibility of a common-view time when the signal pulses can be simultaneously sent from the two OGSs to the Micius satellite is experimental demonstrated in [4-5], which gives a practical foundation of performing space-based MDI-QKD. However, different from the previous works, the MDI-QKD protocol not only has a strict tolerable transmission loss, but also requires the indistinguishability of photons from Alice and Bob. Therefore, some crucial factors such as the choice of experimental parameters [11], the implementations of synchronization [12-13] and the intensity optimization for asymmetric link loss [14] are needed to be further considered in such a large-scale scenario.

[*] **Correspondence:** dongchengfkd@163.com

[†] These authors contributed equally to this work.

In this paper, we present a feasibility assessment of space-based MDI-QKD based on the Micius satellite with the announced experimental parameters. Integrated with the orbital dynamics model and atmosphere channel model, a framework is presented to explore the whole parameters space including orbit height, elevation angle, apertures of transceiver and atmospheric turbulence intensity to give the considerations for improving key rates and subsequently provide a relevant parameter tradeoff for the implementation of space-based MDI-QKD. We further investigate the satellite laser ranging (SLR) technology to achieve synchronization and modify the intensity optimization method so that it can be effectively applied in the dynamic link conditions. Our works can be used as a pathfinder for future quantum communication satellite missions.

The paper is constituted as follows. In Sec. II, the framework for estimating the secure key rate of space-based MDI-QKD is briefly introduced. Then, in Sec. III, we discuss the impacts on the key rate generation from the different considerations in terms of link loss, durations and two-photon interference. The article is ended in Sec. IV with a concluding remark.

## II. PRELIMINARY

We consider an implementation of space-based MDI-QKD between two remote OGSs. As shown in Fig. 1, Alice and Bob are the MDI-QKD photon source that randomly and individually prepares one of four BB84 states using the phase-randomized weak coherent pulse (WCP) sources together with decoy signals [15]. Then they transmit signals to an untrusted third party, Charlie, who is supposed to perform the Bell state measurement (BSM) on a satellite, and announces her measurement results through a public radio channel.

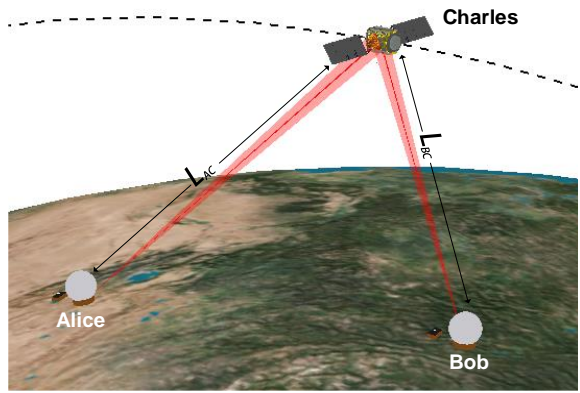

**Fig. 1.** A schematic of space-based MDI-QKD configuration between two remote OGSs (Alice and Bob), where

Charlie hosted on an untrusted satellite. When performing space-based MDI-QKD, the link length of channel between Alice (Bob) and Charles $L_{AC}(L_{BC})$ is asymmetric and time-varied.

In contrast to the terrestrial fiber-based MDI-QKD, as the third party, an untrusted satellite flies along an orbit makes the uplink condition varied in real time. Various adverse effects [16] such as geometric spread, atmospheric turbulence induced fading, background noise, etc., in practice incurs very high loss rates which results in key rate degradation of the space-based MDI-QKD. To ensure an accurate estimation, we here propose a comprehensive integration on the dynamic channel modeling in line with the specific characteristics and scenarios.

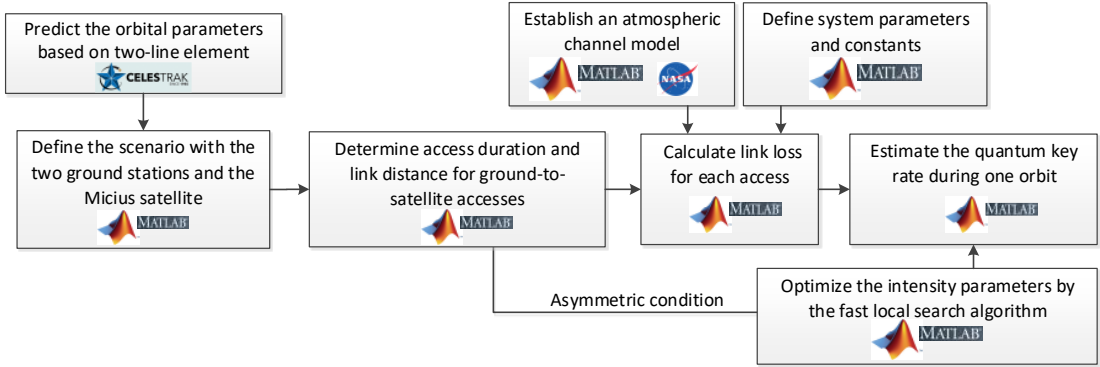

**Fig. 2.** The framework for modeling space-based QKD. First, based on the data file format referred to as two-line element (TLE) [17] sets, an exact orbital element of a specific satellite together with practical geographic locations of OGS can be used as inputs to determine the access duration and link distances. Then, integrated with the atmosphere data provided by the NASA historical cloud statistics [18] and the atmospheric channel calculation model, the link loss for each access can be calculated based on the practical system parameters and experimental constraints. Subsequently, by submitting the values into Eq. (4), the key rate of space-based MDI-QKD per orbit can be estimated. Moreover, the real-time asymmetric channel conditions can also be characterized by our framework to achieve an adaptive parameter optimization.

As illustrated in Fig. 2, a novel framework is proposed to estimate the key rate which is radically different from the simple calculation model proposed in the previous literatures [19-22]. Specifically, the precise orbital model is used to extract the elevation angle $\varphi$ and dual uplink length $L$ for calculating the uplink transmittance. Here, the uplink transmittance $\eta(t)$ at time $t$ for Alice (or Bob) can be given by [23]

$$\eta(t) = \eta_0 \left[1 - \exp\left(-\frac{2R_r^2}{\omega_r(t)^2}\right)\right] \qquad (1)$$

where $\eta_0$ represents the other losses (such as the Rayleigh scattering and absorption, transmitting antennas loss, coupling loss and detection transmittance) and $R_r$ is the radius of the receiving aperture at satellite.

For the uplink, beam width remains smaller than the outer scale of the turbulence during its atmospheric path, and the beam arrives at the receiver with a significant beam wander. Therefore, for a collimated beam, the received long-term beam width $\omega_r(t)$ at time $t$ can be calculated by [24]:

$$\omega_r(t) = L(t) \frac{\lambda}{0.632 R_s \pi} \left[1 + \frac{0.83}{\sin\varphi(t)}\left(\frac{2R_s}{r_0(t)}\right)^{5/3}\right]^{3/5} \qquad (2)$$

Here, $\lambda$ is the laser wavelength and $R_s$ is the radius of the transmitting aperture at the OGS.

Since the uplink path $L(t)$ is a straight line along the real-time elevation angle $\varphi(t)$, the Fried parameter $r_0(t)$ can be expressed as [13]:

$$r_0(t) = \frac{1.1654 \times 10^{-8} \lambda^{1.2} [\sin\varphi(t)]^{0.6}}{\left(\int_0^{Z_{\max}} C_n^2[z(t)]\,\mathrm{d}z\right)^{0.6}} \quad (3)$$

where the $C_n^2[z(t)]$ is the altitude-dependent refractive-index structure constant. The altitude $z(t)$ is shorter by a factor approximately equal to $L(t)\sin\varphi(t)$. For practicality, we set the max altitude of turbulence $Z_{\max} = 20\ km$.

When using the WCPs with decoy states, the secure key rate of MDI-QKD in the asymptotic case is given by [25-27]

$$R(t) \geq P_{11}^Z Y_{11}^Z [1 - H_2(e_{11}^x)] - Q_{\mu\mu}^z f_e(E_{\mu\mu}^z) H_2(E_{\mu\mu}^z) \quad (4)$$

where $Y_{11}^Z$ and $e_{11}^x$ are, respectively, the yield (the conditional probability that Charles declares a successful event) in the rectilinear (Z) basis and the error rate in the diagonal (X) basis, given that both Alice and Bob send single-photon states; $P_{11}^Z$ denotes the probability that Alice and Bob send single-photon states in the Z basis; $\mu$ is the intensity of the signal state; $Q_{\mu\mu}^z$ and $E_{\mu\mu}^z$ denote, respectively, the gain and quantum bit error rate (QBER) in the Z basis, which are simulated here for rate estimation using a known channel transmittance $\eta(t)$; $H_2$ is the binary entropy function.

### III. PERFORMANCE ANALYSIS

Given the realistic input parameters, our proposed framework can be used for presenting a feasibility assessment of space-based MDI-QKD in a specific scenario. In this section, we first explore the whole parameter space to understand under which conditions it may become feasible. The intensity optimization method is employed to promote the key rate generation under the asymmetric channel situation.

#### A. The feasibility of space-based MDI-QKD based on Micius satellite

We use the practical parameters to estimate the key rate achievable for a space-based MDI-QKD under our framework. Here, we choose the existing OGSs (i.e., Delingha and Lijiang) to be as the two remote users, and the Micius satellite as the measurement node, i.e., Charlie. The parameters from the Micius experiments [2] that influence the key rate generation, together with their reference values are summarized in Table I.

TABLE I List of practical device parameters for numerical simulations

| | | |
|---|---|---|
| $\theta$ | divergence angle at the transmitter | 14 μrad |
| $R_s$ | radius of the transmitter aperture | 6.5 cm |
| $R_r$ | radius of the receiving aperture | 15 cm |
| $\lambda$ | signal wavelength | 780 nm |
| $C_0$ | refraction structure parameter at ground | $1.7 \times 10^{-14}\ m^{-2/3}$ |
| $\varphi$ | lowest elevation angle in degree | 10 |
| $\eta_o$ | optical transmittance in dB | 1.5 |
| $\eta_T$ | transmitting antennas loss in dB | 1.5 |
| $\eta_c$ | coupling loss in dB | 5.9 |
| $\eta_d$ | detection efficiency loss in dB | 3 |
| $e_0$ | error probability of dark counts | 0.5 |
| $e_d$ | error probability of optical misalignment | 0.015 |
| $f_e$ | error-correction efficiency | 1.16 |
| $Y_0$ | fixed background rate | $3 \times 10^{-6}$ |

The experiment of ground-to-satellite quantum teleportation through an uplink from Nagari station to the Micius satellite is modeled to validate our atmospheric channel model. The results for uplink characterization are shown in Fig. 3, consistent with the previous Micius experiments.

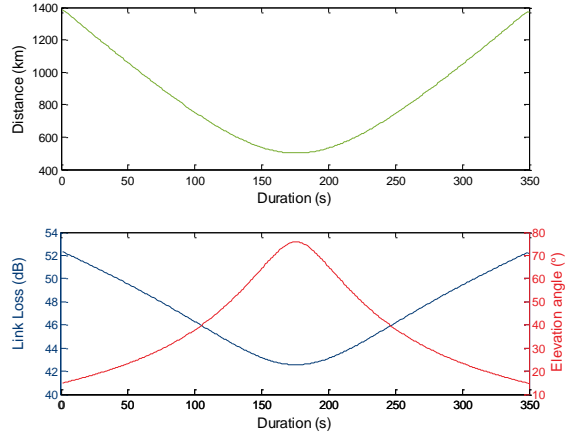

**Fig. 3.** Distance and elevation angle from the Nagari station to the Micius satellite and the attenuation during one orbit. The measured ground-to-satellite channel loss using a strong reference laser as a function of time. The highest loss is about 52.3 dB at a distance of 1,385 km (when the satellite is at an angle of 15°). The lowest loss is about 42.5 dB at a distance of around 501 km (when the satellite is at an angle of 75.9°). Since the [3] did not provide a specific experimental time, the orbital elements for our model is obtained by given the TLE sets in Sep 25 12:36:04 2016 UTCG, which may exist an orbital drift.

Furthermore, the communication duration and the link attenuation are shown in Fig. 4, which are then used to estimate the expected key rate achievable for a link in our assumptions. There is a total of 278 seconds available for establishing the dual uplinks per orbit. However, the link loss for the overall system on the right side of Fig. 4 ranges from 94.0 dB to 100.2 dB. It is somewhat of a concern for the implementation of a MDI-QKD since the previous works on fixed-attenuation channels clearly demonstrate the non-viability of such a protocol in the high-loss regime (The maximum tolerable loss of MDI-QKD [28] is no more than 65 dB). Nonetheless, it illustrated that the space-based MDI-QKD with dual OGSs is difficult to achieve, at least under the current parameters of the Micius satellite.

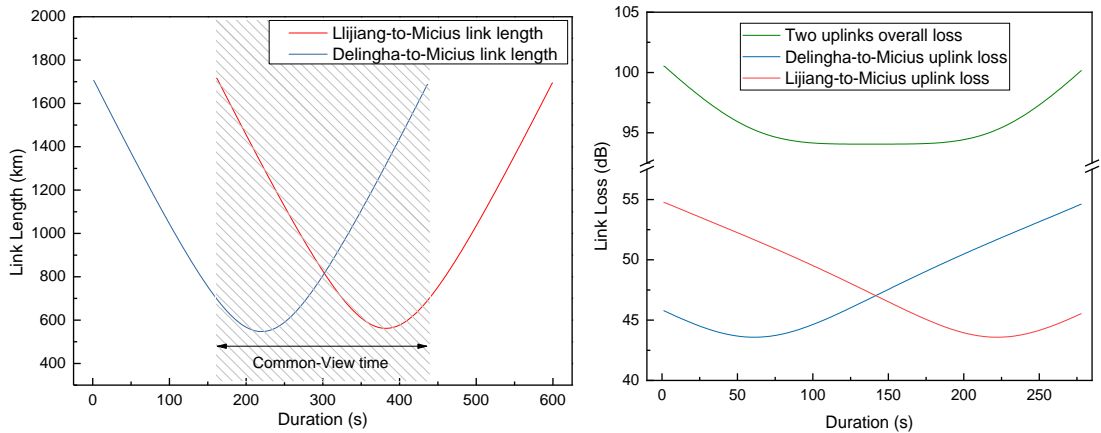

**Fig. 4.** Distance and Link loss from the two stations to the Micius during one orbit. a. The trajectory of the Micius satellite simulated from the Delingha and Lijiang over one orbit with a duration of 278s. b. The highest loss is about 100.2 dB at a distance of 2,428 km, and the lowest loss is about 94 dB at a distance of 1,611 km.

**B. Link loss considerations**

As discussed previously, the main factors affecting the loss of uplink between the OGS and the satellite are the beam wander and the diffraction loss, determined by the turbulence intensities $C_n^2$ and the apertures of the transceiver (i.e., the radius of receiving telescope at satellite $R_r$ and the radius of transmitting telescope located in OGS $R_s$ ). The beam wander is characterized by the divergence of the beam. As for the diffraction loss characterized by radius of receiving telescope at satellite $R_r$, larger radius is considered more attractive while flying large telescopes in space is complex and costly. Therefore, the first big challenge to achieve the space-based MDI-QKD is a relevant parameter tradeoff for its contribution to the link efficiency.

In order to show the benefits through the optimization of parameters more intuitively, we explore the whole parameters space which is compatible with the requirements of the space industry. The variation of link loss is presented in Fig.5. In Fig. 5(a), once the radius of the receiving telescope mounted on the satellite and the transmitting telescope located in OGS is up to 1.2 m and 0.75 m, respectively, the total loss of two uplinks can be significantly decreased to 55 dB, which is within the acceptable link loss range of performing MDI-QKD. The example shows the optimizing aperture parameters to achieve a possible feasibly of the satellite-based MDI-QKD. Therefore, we use such the settings of the apertures that allows the a possible of the satellite-based MDI-QKD in the remaining text. Based on the optimized parameters, we investigated the further deterioration of link loss under different turbulence intensities to get a realistic significance of our assessment. As shown in Fig 5. (b), the lower blue and red curves correspond to the uplinks in the absence of atmospheric turbulence, whereas the shaded bands correspond to the free-space channel with the atmospheric turbulence intensities range of $1.7 \times 10^{-15} \text{ m}^{-2/3} \leq C_n^2 \leq 1.7 \times 10^{-13} \text{ m}^{-2/3}$ . From the figure it is evident that the total link loss under the improved parameters can be smaller than 65 dB during the 278 s of the orbit reserved for space-based MDI-QKD, which indicates that the two-quantum uplinks for performing MDI-QKD are expected to be established in the days with favorable meteorological conditions. Note that, though the link loss of an uplink could be reduced using a larger lens, strong accelerations along the optical axis expected during launch are a serious concern, as is radiation damage of the optics. For this reason, a more complicated tradeoff will be major part of the practical engineering design.

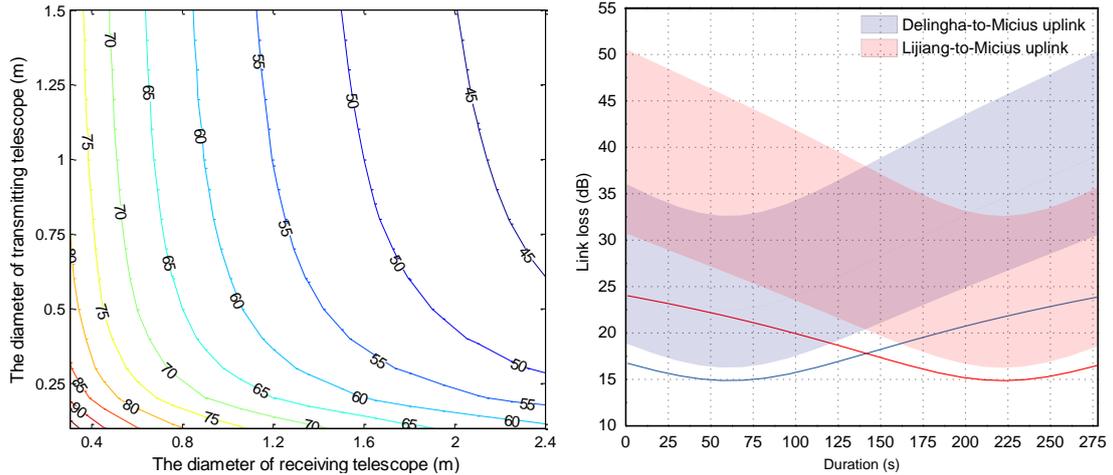

**Fig. 5.** Different factors induced link loss (in dB). **(a)** Total link loss at the shortest overall link length as a function of the apertures of the transceiver. To be compatible with the requirements of the space industry, we take the Hubble Telescope [29], one of the largest optical telescopes in Low Earth orbit (LEO), as an example, with an aperture of 2.4m. Meanwhile, the max aperture of transmitting telescope is chosen as 1.5m (i.e., the observatory located in Matera, Italy [30]). For simplicity, other parameters remain constant. **(b)** The variations of two uplink loss under different intensity of atmospheric turbulence during one orbit, where the aperture of the receiving and transmitting telescope is chosen as a moderate value, i.e., 1.2m and 1m, respectively.

### C. Duration considerations

The Micius satellite which flies along a sun-synchronous orbit cannot be within the common visibility area of the dual OGSs in all time, which drives the choice of orbit parameters by the desire to maximize the uplinks duration. Increasing access duration seems to be beneficial for generating more quantum key bits, while it may also introduce more adverse impacts on the key generation. For instance, increasing the satellite altitude where the MDI-QKD is performed can effectively increase the duration time, but can also introduce more losses because of the increment of received beam width. Alternatively, another approach to increase the link duration is expanding the range of the elevation angle $\phi$, but the path will be longer by a factor approximately equal to $1/\sin(\phi)$, leading to a smaller Fried parameter. Therefore, a tradeoff between link loss and access durations needs to be further discussed.

The results for an orbit of 400-600 km and a minimum elevation angle range of 5-15 degrees are shown in Fig. 6(a). As expected, the link durations over which the two OGSs are simultaneously visible increases with the orbit ascend and elevation angle range widens. However, we point out that the total secure key rate during one orbit is not obvious improved with respect to the magnitude of the increase with link durations as seen in Fig. 6(b). This is because of that, despite the link duration gets extended, the asymmetry of the dual uplink during those increased times is worse. Especially at a lower elevation angle, the link will have to deal with more complicated situations (e.g. urban area), just as the case of the horizontal free-space channel. In addition, requirements to the attitude control system as well as the alignment considerations and power budget are also relaxed in our model, which will further affect the performance of space-based MDI-QKD in practice. What's more, the key rate of MDI-QKD heavily relies on the level of symmetry between the two channels since it requires highly balanced intensities arriving at Charles on the two arms in the X basis for good HOM interference [31].

Therefore, unlike the space-based QKD (whether by uplink or downlink), a better visibility of the two-photon interference is more urgently required for the space-based MDI-QKD rather than increasing the duration. Details will be discussed in the next section.

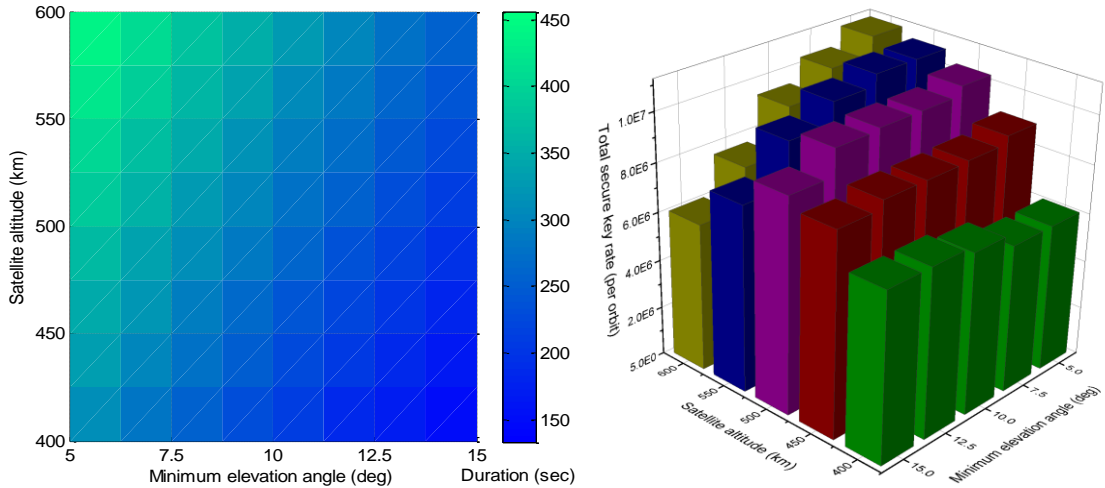

**Fig. 6.** Total access duration and secret key rate (per pulse) during one orbit. **(a)** Link durations under different height and minimum elevation angle. The minimum orbit altitude of around 400 km is the height of the ISS orbit [32]). **(b)** Total secret key rate during one orbit. The sending intensities are $\mu = 0.5$, $\nu = 0.1$ and $\omega = 0$, respectively. Other parameters are consistent with the initial values of parameters in the above works. For simplicity, we assume that the total key rate is the sum of the estimated key rate per second during one orbit. Note that, the key rate is calculated in the asymptotic case, and the finite key rate with the statistical fluctuations should also be analyzed for practicality.

### D. Two-photon interference considerations

In the previous section we can see that MDI-QKD requires the indistinguishability of photons from Alice and Bob, to maintain good HOM interference visibility in X basis. In this perspective, there are some dimensions need to consider: spectrum, polarization and timing [33]. To solve these challenges, the feedback temperature control units for the frequency-locked lasers [34, 35] are employed to match the spectral mode, and the polarization stabilization system is also always essential in the polarization-encoding scheme.

Recently, Pan's team experimentally demonstrated a 19.2-km free-space MDI-QKD with asymmetric and unstable channels [36]. A hydrogen cyanide molecule (HCN) cell is employed to calibrate the wavelength of two distributed feedback laser diodes (LD), which ensure that a frequency offset between the lasers remains below 10 MHz. However, the Doppler shift makes the frequency calibration more complicated in practical satellite-based MDI-QKD scheme. The Fig. 7(a) clearly reflect the impact of this constraint, which shows a shift of about +/-10 GHz in frequency caused by the Doppler effect in the two uplinks during the duration. Due to the relative motion of the satellite and the two OGSs, a flexible calibration is needed to dynamic adjust the frequency of the signals as the satellite approaching and departing. In addition, the receiver must accommodate a frequency offset of about the order of magnitude of GHz to implement a frequency mode matching when performing HOM interference, which seems to be hindered to realize at present. Interestingly, Figure 7(b) shows that an improvement when choosing a longer wavelength laser (1550nm) has up to 10 GHz lower frequency offset than that of the 780 nm laser, will would be a one of the options for future space-based MDI-QKD.

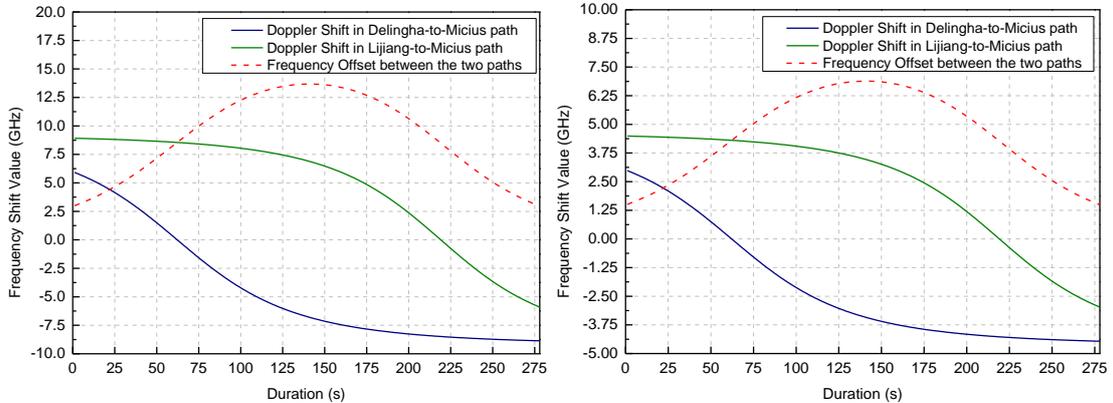

**Fig. 7.** The variation of the frequency shift in the space-based MDI-QKD using different wavelength lasers. (a) using a 780 nm laser. (b) using a 1550 nm laser. A shift of about +/-10 GHz in frequency is caused by the Doppler effect in the two uplinks during the duration, which results in a dynamic frequency offset between two lasers.

On the other the hand, the arrival time of signal pulses is different as they depend on the asymmetry of uplinks under the MDI-QKD setting. In the ground demonstration experiment, a portion of the arriving photons is directly measured in superconducting nanowire single photon detectors (SNSPD) in the measurement node, where arriving time difference of the clock can be estimated, results in a 32 ps of standard deviation of the arrival time difference in a long-term [36]. However, for the two asymmetric uplinks, arriving time difference is dynamic. To precisely match the timing of pulses, the arrival time of signal pulses can be estimated by measuring the temporal separation of two consecutive satellite laser ranging (SLR) pulses at the OGSs after the satellite retroreflection [30]. The synchronization for arrival time can be achieved by adding a difference of the sending time between Alice and Bob. As shown in Fig. 8, the aperiodic arrival time of qubits can be determined by the satellite radial velocity. By selecting the counts obtained within a given temporal window around the arrival time, it is possible to discriminate the qubits coming from the satellite from the background radiation and dark counts [37].

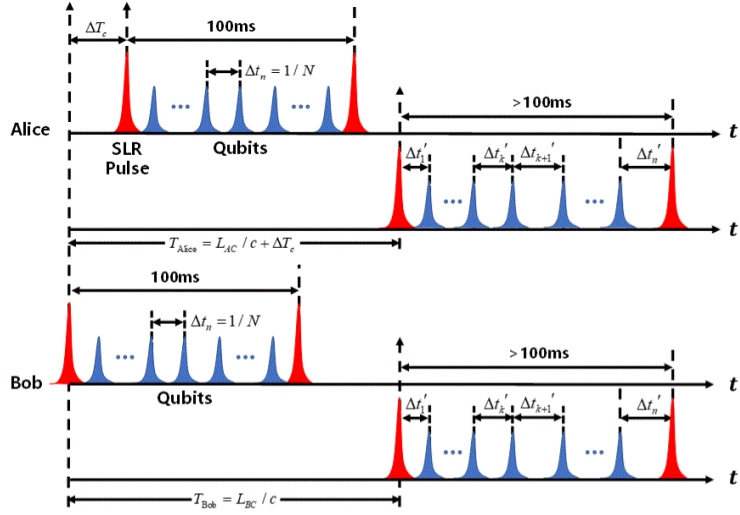

**Fig. 8.** The schematic of the synchronization method for the asymmetric MDI-QKD setting. The synchronization was performed by using the bright SLR pulses. Although the arrival time of the pulses caused by the Doppler effect is aperiodic, the synchronization can be achieved once the sending time of pulses between Alice and Bob is well matched by adding a difference of the sending time between Alice and Bob $\Delta T_c = |L_{AC} - L_{BC}|/c$, where $c$ is the light speed.

Last but most important for space-based MDI-QKD, using same intensities of the WCP sources of Alice and Bob (hence different intensities arriving at Charles after the channels' attenuation) will also result in high QBER of $E_{nm}^x$ in X basis when Alice sends $n$-photon pulse, Bob sends $m$-photon pulse, and consequently poor estimation of the single photon error rate $e_x^{11}$ when two channels are asymmetric. The related the parameters are listed in Table II, which are taken to the equations [38] used to calculate QBER. As shown in Fig. 9, the estimated single photon error rate $e_x^{11}$ in the X basis induced by the asymmetric link losses is lowest when the link losses of two channels are symmetric, since intensities arriving at Charles' beam-splitter are almost equal.

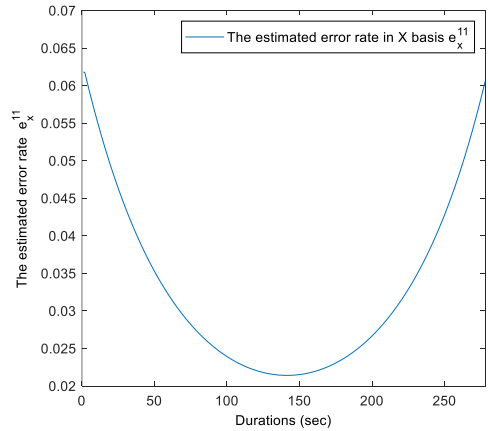

**Fig. 9.** The variation of the single photon estimated error rate in X basis in the space-based MDI-QKD using WCP sources. Here, we use the infinite decoy method to estimate the QBER, and a higher QBER can be expected when considering the finite-size case [39].

Therefore, to improve the performance of space-based MDI-QKD, we consider the intensity optimization (IO) [28] that allows users to independently choose their intensities of the signal state and the decoy state to compensate for the different channel losses. However, considering the dynamic asymmetric channel condition, a tradeoff between the operation time and the accuracy of optimal solution should be taken into account. Here, we divide the durations into different time slots (5s, 10s, 15s, 20s and 25s) and perform the IO for per time slot. The results of total key rates per orbit (with a finite data set $N=10^{14}$ every second) are shown in Fig. 10. Compared with the key rate using a fixed parameters set of μ=0.5, ν=0.1, the maximum total key rate in MDI-QKD with the optimal intensities is almost two orders of magnitude higher than that without the IO, increasing to $1.9508\times10^8$ bits during one orbit. In particular, such a way to perform the IO based on the averaged link length in each time slot, indeed, could incur a coarse optimal key rate compared to that per second, while whose nature of course can sure a faster response for dynamic links. It is most likely to be implemented in practical space-based MDI-QKD.

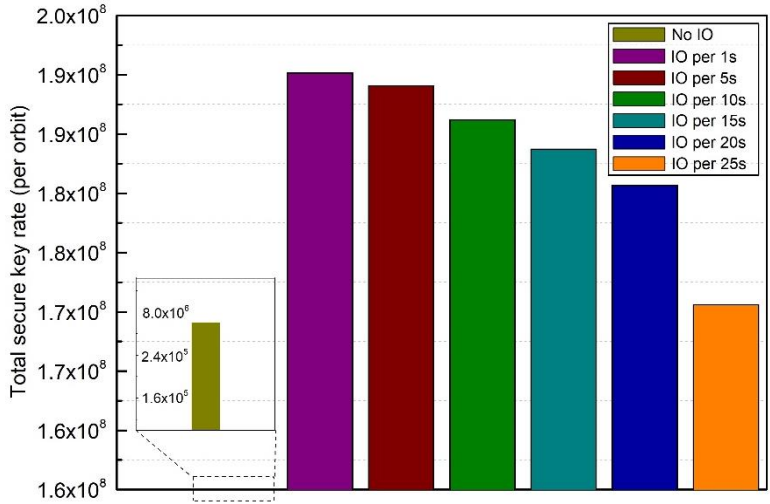

**Fig. 10.** Total key rate comparison with the different optimization methods. From left to right, bars represent the total secret key rates during one orbit without using the intensity optimization method (dark yellow bar), with intensity optimization performed in different size of time slots: per second (purple bar), per 5 seconds (wine bar), per 10 seconds (green bar), per 15 seconds (dark cyan bar), per 20 seconds (dark blue bar) and per 25 seconds (orange bar), respectively.

## IV. CONCLUSION

In this work we have explored a feasibility assessment of the space-based MDI-QKD for two ground stations via an untrusted satellite. Integrated with the orbital dynamics model and atmospheric channel model, a framework was proposed to estimate the total link budget for an orbital scenario where the Micius satellite passes the two practical OGSs (located in Delingha and Lijiang). For this we used conservative estimates of the uplink loss due to the diffraction and beam wander. Subsequently exploring the parameters that are associated with orbit elements and optical systems, we gave the considerations to improve key rates of the MDI paradigm. Furthermore, we have investigated the influences of asymmetric channel condition on the indistinguishability of photons. We divided durations into time slots and performed an intensity optimization per time slot. Compared to the real-time intensity optimization, the modified method showed a same order of magnitude in the key rate generation, which is more likely to be applied in a practical space-based MDI-QKD. At the same time, we found that a frequency offset of about the order of magnitude of GHz to implement a frequency mode matching when performing HOM interference. Also, the approaches such as use of the SLR technology may be available and could achieve a two-photon interference for the future. Together, our preliminary assessment could be a guideline for the future quantum satellite missions.

We note that, some physical parameters, such as apertures of the transceiver and diffraction losses, are expected to become relatively more important requirement drivers of an experiment system design. In a real satellite mission, the impact of fine pointing and tracking, the space radiation on the system and the finite key considerations are also crucial. Therefore, from a system engineering perspective, there is a more complex tradeoff in how to choose the parameters to achieve an optimal performance. Future work in this area may explore this issue in more detail, possibly extending the scope to include even the aforehand topology design to establish satellite-based quantum communication network.

**COMPETING INTERESTS**
The authors declare that they have no competing interests.

**FUNDING**
We acknowledge support from the National Natural Science Foundation of China (Grant No.11704412), Key Research and Development Program of Shaanxi (Program No. 2019ZDLGY09-01), Innovative Talents Promotion Plan in Shaanxi Province (Grant No. 2020KJXX-011) and Key Program of National University of Defense Technology (19-QNCXJ-009).

**AUTHOR CONTRIBUTIONS**
Xingyu Wang performed the bulk of this work at the National University of Defense Technology during an exchange stay. Part of Satellite-QKD model was worked by Chen Dong and Xingyu Wang in this work. Shanghong Zhao and supervised Wang at School of Information and Navigation. With thanks to Yong Liu, Xiaowen Liu and Haonan Zhu from our research group who mooted the concept of the space-based QKD network with us.

**ACKNOWLEDGEMENTS**
We thank Wenyuan Wang, and Qin Wang for helpful discussions. Celestrak TLE database of the Micius satellite acknowledges support from the Dr. T.S. Kelso.

**Reference**
1. J. Chen, et al. Sending-or-not-wending with independent lasers: secure twin-field quantum key distribution over 509 km. *Phys. Rev. Lett.,* **124**, 070501 (2020).
2. K. Tamaki, et al. Ground-to-satellite quantum teleportation. *Nature*, **549**, 70–73(2017).
3. S. K. Liao, et al. Satellite-to-ground quantum key distribution. *Nature*, **549**, 43-47 (2017).
4. J. Yin. et al. Space-based entanglement distribution over 1200 kilometers. *Science,* **356**, 1140–1144 (2017).
5. J Yin. et al. Entanglement-based secure quantum cryptography over 1,120 kilometers. *Nature*, **15**, 1-5 (2020).
6. Lo, H.-K., Curty, M. & Qi, B. Measurement-device-independent quantum key distribution. *Phys. Rev. Lett.,* **108**, 130503 (2012).

7. Y. L. Tang, et al. Measurement-device-independent quantum key distribution over 200 km. *Phys. Rev. Lett.,* **113**, 190501 (2014).
8. H. Yin, et al. Measurement-device-independent quantum key distribution over a 404 km optical fiber. *Phys. Rev. Lett.,* **117**, 190501 (2016).
9. R. Bedington,, J.-M. Arrazola, & A. Ling, Progress in satellite quantum key distribution. *NPJ Quantum. Inform.,* **3**, 30(2017).
10. C. Bonato,, A. Tomaello,, V. Deppo, G. Naletto,& P. Villoresi, Feasibility of satellite quantum key distribution. *New journal of physic.* **11**, 045017(2009).
11. J. P. Bourgoin, et al. A comprehensive design and performance analysis of LEO satellite quantum communication. *New Journal of Physic,* **15**, 023006 (2013).
12. D. Dequal, et al. Experimental single-photon exchange along a space link of 7000 km, *Phys. Rev. A,* **93**, 010301(2016).
13. S. P. Neumann, S. Joshi, et al Q3Sat: quantum communications uplink to a 3U CubeSat feasibility & design, *EPJ Quantum Technology,* **5**, 4 (2018).
14. W. Y. Wang, F. H. Xu, and Hoi-Kwong Lo, Asymmetric Protocols for Scalable High-Rate Measurement-Device-Independent Quantum Key Distribution Networks, *Phys. Rev. X*, **9**, 041012 (2019).
15. Tang, Z., Liao, Z., Xu, F., et al. Experimental demonstration of polarization encoding measurement-device-independent quantum key distribution, *Phys. Rev. Lett.,* **112**, 190503 (2013).
16. Oi, D. K. et al. CubeSat quantum communications mission. *EPJ Quantum Technology,* **4**, 6 (2017).
17. Celestrak TLE database of the Micius satellite (Sep 25 12:36:04 2016 UTC). URL http://celestrak.com/satcat/search.php.
18. NASA Cloud statistics database (1 month - terra/modis). URL https://neo.sci.gsfc.nasa.gov/view.php?datasetId=MODAL2_M_AER_OD
19. Hosseinidehaj, N., & Malaney, R. CV-MDI Quantum Key Distribution via Satellite, *Quantum Inf. Comput.*, **17**, 361-379 (2016).
20. Dequal D. et al. Feasibility of satellite-to-ground continuous-variable quantum key distribution. Preprint at https://arxiv.org/abs/2002.02002 (2020).
21. Cai, Y., Lu, X. & Lin Q. Hollow Gaussian beams and their propagation properties, *Optics Letters*, **28**, 1084-1086 (2003).
22. W. T. Liang & R. Z. Jiao. Satellite-based measurement-device-independent quantum key distribution, *New J. Phys.*, **22**, 083074 (2020).
23. Dios, F., Rubio, A., Rodriguez, A. & Comeron, A. Scintillation and beam-wander analysis in an optical ground station-satellite uplink. *Applied Optics*, **43**, 3866–3873(2014).
24. Erik K. et al. Nanobob: a CubeSat mission concept for quantum communication experiments in an uplink configuration, *EPJ Quantum Technology*, **5**, 6(2018).
25. Xu, F., Curty, M., Qi, B. & Lo, H.-K. Practical aspects of measurement-device-independent quantum key distribution. *New J. Phys*, **15**, 113007(2013).
26. Xu, F., Xu, H. & Lo, H.-K. Protocol choice and parameter optimization in decoy-state measurement-device-independent quantum key distribution. *Phys. Rev. A*, **89**, 052333 (2014).
27. Yu, Z.-W., Zhou, Y.-H. & Wang, X.-B. Three-intensity decoy state method for device independent quantum key distribution. *Phys. Rev. A*, **88**, 019901(2013).

28. Zhou, Y.-H., Yu, Z.-W. & Wang, X.-B. Making the decoy-state measurement-device - independent quantum key distribution practically useful. *Phys. Rev. A*, **93**, 042324 (2016).
29. C. F. Prosser, R. C. Kennicutt, Jr, et al. The Hubble Space Telescope Key Project on the Extragalactic Distance Scale. XXII. The Discovery of Cepheids in NGC 1326A. *The Astrophysical Journal*, **525**, 80, (1999).
30. G. Vallone, D. Bacco, D. Dequal, et al. Experimental Satellite Quantum Communications. *Phys. Rev. Lett.,* **115**, 040502 (2015).
31. Z. Y. Tang, Z. F. Liao, F. H. Xu, B. Qi, L. Qian, and H. K. Lo, *Phys. Rev. Lett.*, **112**, 190503 (2014).
32. M. Aguilar et al. The Alpha Magnetic Spectrometer (AMS) on the international space station: Part I - results from the test flight on the space shuttle, *Physics Reports*, **366**, 6, (2002).
33. Eleftherios, M., Ivan, G. J., Joseph, R. B., Bing, Q., Raphael, P., & George, S. Experimental study of Hong–Ou–Mandel interference using independent phase randomized weak coherent states. *J Lightwave Technol.* **36**, 3752 - 3759 (2018).
34. Tang, G., Sun, S., Xu, F., Chen, H., Li, C. & Liang, L. Experimental asymmetric Plug-and-Play Measurement-device-independent quantum key distribution. *Phys. Rev. A* **94**, 032326(2016).
35. Y. Liu, T. Y. Chen, L. J. Wang, H. Liang, G. L. Shentu, J. Wang, et al. *Phys. Rev. Lett.*, **111**, 130502 (2013).
36. Y. Cao, Y. H. Li, K. X. Yang, Y. F. Jiang, S. L. Li, X. L. Hu, M. Abulizi, C. L. Li, W. J. Zhang, Q. C. Sun, W. Y. Liu, X. J., S. K. Liao, J. G. Ren, H. Li, Li. X. You, Z. Wang, J. Yin, C. Y. Lu, X. B. Wang, Q. Zhang, C. Z. Peng & J. W. Pan. Long-distance free-space Measurement-device-independent quantum key distribution. *Phys. Rev. Lett.* **125**, 260503 (2020).
37. E. Anisimova, B. L. Higgins, et al. Mitigating radiation damage of single photon detectors for space applications, *EPJ Quantum Technology,* **4**, 10 (2017).
38. Sun, S., Gao, M., Li. C., & Liang L. Practical decoy-state measurement-device-independent quantum key distribution. *Phys. Rev. A* **94**, 032326(2016).
39. Curty, M., Xu, F., Cui, W., Lim, C. C. W., Tamaki, K & Lo, H.-K. Finite-key analysis for measurement-device-independent quantum key distribution. *Nat. Commun.*, **5**, 5235 (2014).